\documentclass[12pt]{article}

\begin{document}

\sloppy
\begin{flushright}{SIT-HEP/TM-16}
\end{flushright}
\vskip 1.5 truecm
\centerline{\large{\bf Comment on the stability of the Yukawa couplings}}
\centerline{\large{\bf and the cosmological problems of intersecting
brane models}}  
\vskip .75 truecm
\centerline{\bf Tomohiro Matsuda
\footnote{matsuda@sit.ac.jp}}
\vskip .4 truecm
\centerline {\it Laboratory of Physics, Saitama Institute of
 Technology,}
\centerline {\it Fusaiji, Okabe-machi, Saitama 369-0293, 
Japan}
\vskip 1. truecm
\makeatletter
\@addtoreset{equation}{section}
\def\theequation{\thesection.\arabic{equation}}
\makeatother
\vskip 1. truecm

\begin{abstract}
\hspace*{\parindent}
In string theory, stabilization of moduli fields and their cosmological
 implications have been discussed by many authors.
In this paper we do not consider conventional modulus,
nor relative distance between two branes.
We focus our attention to a relative position of three intersecting branes.
Surprisingly, there had been no phenomonological argument on the
 stabilization of such moduli.
We will show that the area of the corresponding 
triangle is not a free parameter, but an effective potential is
generated from conventional loop corrections in 
the low energy effective theory.
Of course, the stabilization does not induce any serious problem, because
one is allowed to modify other parameters of the model to adjust 
the Yukawa couplings.
Then the stabilization puts a constraint that is a
different nature from the ones that have been discussed before.
We also discuss cosmological problems
and show a simple idea that can solve the problem.
\end{abstract}

\newpage
\section{Stability and cosmological problems of intersecting brane models}
\hspace*{\parindent}
In spite of the great success of quantum field theory and classical
Einstein gravity, there is still no consistent unification scenario in which
quantum gravity is successfully included.
Perhaps the most promising scenario in this direction is string
theory, in which consistency of the quantum gravity is ensured by a
requirement of additional dimensions.
Originally the size of extra dimensions was assumed to be as small
as $M_p^{-1}$. 
However, later observations showed that there is no reason
to require such a tiny compactification radius\cite{Extra_1}.
In this respect, what we had seen in the old string theory was a tiny 
part of the whole story.
In the new scenario, the compactification radius (or the fundamental
scale) is an unknown parameter that should be determined by
observations.  
In models with large extra dimensions, 
the observed Planck mass is obtained by the relation $M_p^2=M^{n+2}_{*}V_n$,
where $M_{*}$ and $V_n$ denote the fundamental scale of gravity
and the volume of the $n$-dimensional compact space.
In this scenario the standard model fields are expected to be localized
on a wall-like structure and the graviton propagates in the bulk.
The most natural embedding of this picture in the string theory context 
is realized by a brane construction.
Thus it is quite important to construct models of the brane world 
where the observed fermion spectrum of the standard model is included
in the low energy effective theory.
In this respect, chirality of the fermions and the family replication
are two of the most 
important characteristics of the standard model, which must be included
in the fundamental theory.
Possiblities for fermion chirality in brane construction are
already discussed by many authors.
One of the examples is to locate D3-branes on a
orbifold singularity\cite{chirality-orbifold}.
An alternative is discussed in\cite{chirality-intersection}, where
Dp-branes are put intersecting at non-vanishing angles.
Considering open
strings stretched between them, strings living at the intersecting
points become chiral fermions in four-dimensional effective
Lagrangian. 
Phenomonological aspects of intersecting brane world are discussed
in ref.\cite{Int-brane-world} and many
authors\cite{Int-brane-follow}.

In this paper we discuss stabilization and cosmological aspects of the 
Yukawa couplings in the intersecting brane models.
Here we do not pretend to make it clear the whole story of the
cosmological evolution, nor to discuss all the possible 
problems, but simply focus on some cosmological criteria that will 
become important in generic situations.
Of course, we know that sometimes cosmology of the models for the
braneworld (or models with large extra dimensions) seems quite
peculiar.\footnote{Constructing successful models for  
inflation with a low fundamental scale is still an interesting
problem\cite{low_inflation, matsuda_nontach}.
Baryogenesis and inflation in models with a low fundamental scale are
discussed in \cite{low_baryo, low_AD, ADafterThermal}.
We think constructing models of particle cosmology with large
extra dimensions is very important since future cosmological
observations would determine the fundamental
scale of the underlying theory.}
In any case, we know historically that the characteristic features of
phenomonological models are revealed by discussing their cosmological
problems. 
Then it will be quite natural to consider a question,
``What is the characteristic problem of the intersecting brane models
that would be induced by cosmology?''
In this paper we will discuss one of the possible answers for the above
question. 

In the models of the intersecting branes, several kinds of moduli fields are
expected to exist at low energy effective theory. 
Some of them might be fixed by the natural mechanisms and integrated away
from the effective theory,
while others might 
not.\footnote{Shift of a gauge coupling during inflation is
discussed in \cite{matsuda_moduli}. }
In this paper we focus our attention to the Yukawa couplings, which come
from the instantons of triangles bounded by three branes.
It was noticed in ref.\cite{Int-brane-world, yukawa} that Yukawa
couplings among three fields living at brane intersections will arise
from the calculation of worldsheet instantons involving three boundary
conditions. 

As we will see in the next section, surface area of a triangle
depends also on the volume and the structure of the compactification,
which could be fixed by some known mechanisms.
Thus we consider in this paper the most optimistic scenario in
which the geometry of the compactified space is already fixed and does not
induce another problem.
In this case the ambiguity of the area of the triangle appears only through the
shift of the relative position of the three branes, which cannot be fixed by
the conventional mechanism for the moduli stabilization.

For the distance between two branes, it is known that an effective
potential is generated when supersymmetry is broken.
This potential is a simple perturbative effect of the string, but cannot
stabilize the relative position of {\bf intersecting} branes, because
the distance is already $0$ for such branes.
On the other hand, it was noted in ref.\cite{yukawa}, that in the context of
the intersecting brane world, Yukawa couplings are generated by the worldsheet
instantons involving three different boundary conditions and three
different intersections.
In this paper we will discuss that the effective
potentials generated  
through conventional fermion loop corrections stabilize the Yukawa couplings.
The stabilization of the Yukawa couplings is good news for the
intersecting brane models, although it forces some modifications to
the previous models. 
In this paper, however, we also point out that the cosmological problems
might remain even if one considers the above mechanism for the
stabilization.\footnote{Of course the situation is much better than that
without any mechanism for the stabilization.}
One reason is that generic models of intersecting brane world
contain at least three triangles of intersecting branes that correspond
to three generations, 
which might lead to three degenerated vacua where each triangle 
shrinks.
Then it is natural to worry about the problem of 
cosmological domain walls that interpolate between two of the three
degenerated vacua. 
Even if the fluctuations of the corresponding massless modes are
suppressed, the allowed initial
condition for the brane configuration is quite restricted.
One way to avoid these problems is to include  at least one ``large''
correction to the effective 
potential that involves the areas of the triangles. 
Such a potential will stabilizes the brane configuration during cosmological 
evolution of the early Universe.
Unfortunately, in this case one should have to worry about the significant
correction that might affect the effective Yukawa couplings in the
standard model. 
At present, it seems impossible to include such corrections within the
setups of the conventional intersecting brane models.
Another way is to assume that the translational invariance is explicitly
broken at the beginnings and there is no freedom for the brane positions
to be shifted. 
In this case, one must answer why branes are fixed at the
present positions.
Explicit breaking of the reparametrization invariance would be possible, 
but it requires significant modification of the original scenario.
From the discussions above, it seems rather difficult to construct a
model where the present brane configuration is ensured by introducing
another mechanisms for the stabilization.
Alternatively, one can assume that the three generations are not equal but
slightly different in their geometrical settings, so that the
cosmological domain walls become unstable even if they are produced.
The last example seems to be the most realistic.
For example, one may assume ``tiny warping'' that resolves the
degeneracy of the three vacua and destabilizes the domain wall.
In this case, our criteria will merely put a lower bound for the warping
factor. 
Let us explain the situation in more detail.
The areas of the triangles depend
on the parameter $A$ in 
eq.(\ref{A-form}), which 
represents the K\"ahler structure of the torus.
If the parameter $A$ is slightly warped so that it depends on the generation,
it induces energy difference in (\ref{effective}).
In this case, our criterion for the unstable domain walls puts a lower
bound on the magnitude of the 
energy difference between different vacua.
For $\sigma \geq (10^5 GeV)^3$, the required energy difference
$\epsilon$ is\cite{vilenkin, wall_matsuda} 
\begin{equation}
\label{decay}
\epsilon > \frac{\sigma^{2}}{M_{p}^{2}}
\end{equation}
where $\sigma$ is the tension of the domain wall. 
When the tension of the domain wall is smaller than $(10^5 GeV)^3$, one
should consider another bound from the 
nucleosynthesis\cite{Abel}, 
\begin{equation}
\label{nucle}
\epsilon > \lambda \sigma M_{EW}^2/M_p,
\end{equation}
where $\lambda \geq 10^{-7}$ is required, and $M_{EW}$ is the
electroweak scale. 
In the present model, the tension of the domain wall is about
$\sigma\simeq M_{EW}^3$ and the energy difference 
$\epsilon$ is induced by the shift of the K\"ahler structure $A$, if $A$
is warped.\footnote{To be more precise, here the K\"ahler structure $A$
is supposed to distinguish the generation.
This can be realized by introducing additional breaking of the flavor
symmetry, or by assuming that the positions of the triangles are weakly
fixed on the warped manifold.}
Then it is straightforward to calculate the required bound for the warp
factor from eq.(\ref{nucle}) and (\ref{effective}).\footnote{Here the energy
difference is induced by the light fermions whose Yukawa couplings depend
explicitly on $A$.
Since the area of the heaviest fermion vanishes in each vacuum, the
Yukawa coupling (\ref{yukawa_formula}) of the heaviest fermion can not
depend on $A$.} 
At present one cannot remove the theoretical uncertainties 
concerning the prefactors and numerical factors in the exponents, but
the result suggests that the tiny shift in the K\"ahler structure $A$
is enough to remove the unwanted domain walls. 
Since the required warping is tiny, one can expect that the small
pertubvation induced by the additional flavor-violating components might
solve the problem, which is ignored so far.
Of course, the problem is not confined to the issue of the stability of
the Yukawa couplings, but should be solved including the stabilization
of the whole moduli fields, which should be discussed in the forthcoming
papers.

One might also think that the relative position
of the intersecting three branes could be fixed without introducing any
mechanism for the stabilization, if the homogeneous initial
condition is achieved during inflation.
In this case, however, it is difficult to construct models where the
fluctuations of the positions of the branes are safely 
suppressed throughout inflation and reheating.\footnote{In this sense, brane
collision cannot produce intersecting braneworld.}
In models where the electroweak symmetry is spontaneously broken by
radiative corrections, there is a lower bound for the largest Yukawa
coupling, which suggests that the initial condition
(and its fluctuation) must be finely tuned to be within the restricted
area.
Obviously, phenomonological bound is so tight that the unnatural
fine tuning is required, even if our mechanism for the stabilization works.
Otherwise, the electroweak symmetry breakdown does not start because
none of the fermions develops $O(1)$ Yukawa coupling.

In any case, changes are required for the present models of
the intersecting brane world.
Of course, the situation is much better than the models where
stabilization of the Yukawa couplings is ignored.
As we have discussed above, once the effect of the stabilization is included,
a small warp factor may solve the problem of
the initial brane condition, because it can destabilize the unwanted domain
walls.

\section{1-loop potential and vacuum degeneracy}
\hspace*{\parindent}
In this section  we consider the simplest example
in ref.\cite{Int-brane-world, yukawa}.
What we would like to see is the Yukawa couplings in the quark sector.
The Yukawa coupling among two chiral fermions and one Higgs boson
cannot appear from perturbative effects of the string theory, but
induced by worldsheet instanton corrections of the corresponding
triangle that has three boundaries of the intersecting branes and three
vertices where matter fields live.

Here we consider the simplest case and derive the
expression for Yukawa couplings.
When computing a sum of worldsheet instantons, the simplest example
comes from D-branes wrapping 1-cycles in a $T^2$, where branes are
intersecting at one angle.
Here we associate each brane to complex number $z_{\alpha},
(\alpha=a,b,c)$, 
\begin{eqnarray}
z_a&=&R\times (n_a+\tau m_a)\times x_a\nonumber\\
z_b&=&R\times (n_b+\tau m_b)\times x_b\nonumber\\
z_c&=&R\times (n_c+\tau m_c)\times x_c.
\end{eqnarray}
Here $(n_{\alpha},m_{\alpha})\in {\bf Z}^2$ denote the 1-cycle the brane
$\alpha$ wraps on $T^2$ and
$x_{\alpha} \in {\bf R}$ is an arbitrary number.
$\tau$ is the complex structure of the torus.
These branes are given by a straight line in ${\bf C}$. 
The triangle corresponding to a Yukawa coupling must involve three branes,
which has the form $(z_a,z_b,z_c)$ with $z_z+z_b+z_c=0$.
The solution is 
\begin{equation}
x_{\alpha}=I_{\beta\gamma}x/d,
\end{equation}
where $x=x_0+l, x_0 \in {\bf R}, l \in {\bf Z}$ and
$d=g.c.d.(I_{ab},I_{bc},I_{ca})$.
Here $I_{\beta\gamma}$ stands for the intersection number of branes
$\beta$ and $\gamma$.
Indexing the intersection points, one can obtain a simple expression for 
$x_0$\footnote{See ref.\cite{yukawa} for more detail.},
\begin{equation}
x_0(i,j,k)=\frac{i}{I_{ab}}+\frac{j}{I_{ca}}+\frac{k}{I_{bc}}
+\frac{I_{ab}\epsilon_c + I_{ca}\epsilon_b +I_{bc}
\epsilon_a}{I_{ab}I_{bc}I_{ca}},
\end{equation}
where the parameter $\epsilon_\alpha$ correspond to shifting the
positions of the three branes.
Using this solution, one can compute the areas of the triangles whose
vertices lie on the triplet of intersections $(i,j,k)$,
\begin{equation}
\label{A-form}
A_{ijk}(l)=\frac{1}{2}(2\pi)^2 A|I_{ab}I_{bc}I_{ca}|\left(
x_0(i,j,k)+l\right)^2
\end{equation}
where $A$ represents the K\"ahler structure of the torus.
The corresponding Yukawa coupling is given by 
\begin{equation}
\label{yukawa_formula}
Y_{ijk}\sim \sigma_{abc}\sum_{l\in{\bf Z}}
exp\left(-\frac{A_{ijk}(l)}{2\pi\alpha'}\right),
\end{equation}
where $\sigma_{ijk}=sign(I_{ab}I_{bc}I_{ca})$ is a real phase.

Now our question is how one can determine the areas of the triangles.
A perturbative force between branes can
produce potential for the distance between two branes.
However, it is obvious that this force cannot affect the area of a
triangle when branes are {\bf intersecting}.
On the other hand, one can see from eq.(\ref{yukawa_formula}) that
almost all the parameters are determined if the windings of the
branes and the structure of the manifold are fixed by some mechanisms.
The only ambiguity that might remain at low energy effective theory is
one parameter of three $\epsilon_\alpha$, which corresponds to shifting
the relative brane position. 
For the area of a triangle, only one of the three parameters
$\epsilon_\alpha$ is independent.

An effective potential for the area of a triangle is obtained by
considering a well-known 1-loop correction from fermion loops
of the form\cite{Kolb-turner}
\begin{equation}
\label{effective}
\Delta V(\phi_c)=-\frac{3}{64\pi^2} Y_{ijk}^4 \phi_c^4 
ln\left(\frac{\phi_c^2}{\mu^2}\right),
\end{equation}
where $\phi_c$ denotes the classical field.
From eq.(\ref{effective}) and (\ref{yukawa_formula}), one can easily 
see that the 1-loop correction stabilizes the area of the triangle.

Because of the exponential form of the potential, intersecting branes are
stabilized when one of the areas of the three triangles vanishes.
In general, models for the intersecting branes
are designed so that the three triangles cannot
shrink simultaneously to a point, so that they satisfy the
phenomonological requirements.
Thus it is straightforward to construct the models in which
one of the 
three Yukawa 
couplings becomes large, while others remain (hierarchically)
small.\footnote{Note that the triangle of the largest Yukawa coupling always
shrinks to a point.
As a result, the areas of the triangles are no longer the free
parameters, but are always determined once the geometrical settings are
fixed.}

\section{Conclusions and Discussions}
\hspace*{\parindent}
Stabilization of the moduli fields and their cosmological implications 
are quite important in any phenomenological models of string theory.
In this paper we examined a peculiar cosmological problem of the moduli
field in models of the intersecting brane world.
We focused our attention to a relative position of three branes and
 discussed its cosmological problems.
The relative position of the intersecting three branes is not determined by
conventional perturbative effects.
To obtain an effective potential for the corresponding moduli,
one should consider string effects that have more than two
 boundaries. 
In models of the intersecting brane world, instanton effects
 that have three 
 boundaries play an important role in determining Yukawa couplings.
We considered a simple model and discussed how the relative position
among three intersecting branes  
appears in the effective potential.
We showed that the effective potential that contains the area of the
triangle is 
generated by the 1-loop corrections of fermions, which involve the Yukawa
couplings.
Cosmological aspects of the model are also discussed in this context.
The constraints on the models become slightly stronger if one includes the
stabilization of the Yukawa couplings, since 
the area of the triangle is no longer a free parameter of the model.
However, once the stabilization is included, there appears a chance to solve
the cosmological problem of the initial brane configuration.
As we have discussed, a tiny warp factor can induce the
energy difference between degenerated vacua, which removes the
cosmological domain walls.

\section{Acknowledgment}
We wish to thank K.Shima for encouragement, and our colleagues in
Tokyo University for their kind hospitality.

\end{document}